\documentstyle[aps,prd,preprint]{revtex}
\newcommand{\covar}{{\not \!\! D}}
\newcommand{\tr}{ \:{\rm{tr}} \:}
\newcommand{\Seff}{{\rm{S}}_{eff}}
\newcommand{\sech}{{\rm{sech}}}
\newcommand{\csch}{{\rm{csch}}}
\begin{document}
\draft
\title{On the QED Effective Action in Time Dependent\\Electric
  Backgrounds}

\author{Gerald Dunne \thanks{dunne@hep.phys.uconn.edu} and
  Theodore Hall  \thanks{tedhall@phys.uconn.edu} }
\address{Physics Department, University of Connecticut,
  Storrs, CT, 06269 USA}

\maketitle

\begin{abstract} We apply the resolvent technique to the
computation of the QED effective action in time dependent electric
field backgrounds. The effective action has both real and imaginary
parts, and the imaginary part is related to the pair production
probability in such a background. The resolvent technique has
been applied previously to spatially inhomogeneous magnetic
backgrounds, for which the effective action is real. We explain
how dispersion relations connect these two cases, the magnetic
case which is essentially perturbative in nature, and the
electric case where the imaginary part is nonperturbative.
Finally, we use a uniform semiclassical approximation to find
an expression for very general time dependence for the
background field. This expression is remarkably similar in
form to Schwinger's classic result for the constant electric
background.
\end{abstract}
\pacs{}

\section{Introduction}
The effective action is an important tool in quantum
electrodynamics (QED), and quantum field theory in
general. For example, for fermions in a static magnetic
background, the effective action yields (minus) the effective
energy of the fermions in that background; while for fermions
in an electric background, the effective
action is complex and the imaginary part gives (half) the
probability for fermion-antifermion pair production
\cite{schwinger,stone}.

The computation of rates for pair-production from the vacuum was
initiated by Schwinger \cite{schwinger} who studied the constant field
case and found that the rate is (exponentially) extremely
small. Brezin and Itzykson \cite{brezin} studied the more realistic
case for alternating fields $\vec{E}(t)=(E \sin(\omega_0 t),0,0)$ but
found negligible frequency dependence and still an unobservably low
rate for realistic electric fields. Narozhny\u{i} and Nikishov
\cite{naro} obtained an expression for both the spinor QED and scalar
QED effective action, as an integral over 3-momentum, for a time
dependent field $\vec{E}(t)=(E \sech^2\left({t\over
\tau}\right),0,0)$. Their approach was based on the well-known exact
solvability of the Dirac and Klein-Gordon equations for such a
background. This solvable case
has also featured in the strong-field analysis of Cornwall and
Tiktopoulos \cite{cornwall}, the group-theoretic semiclassical
approach of Balantekin {\it et al.} \cite{balantekin88,balantekin91}, the
proper-time method of Chodos \cite{chodos},
and the $S$-matrix work of Gavrilov and Gitman \cite{gavrilov}. Recent
experimental work involving the SLAC accelerator and intense lasers
has given renewed impetus to this subject, providing tantalizing hints
that the critical fields required for direct vacuum pair production
may be within reach \cite{burke,melissinos}.

In this paper we make several new contributions to this body of
work. First, using the resolvent approach we present an expression for
the exact effective action in the time-dependent background
$\vec{E}(t)=(E \sech^2 \left( {t\over \tau} \right),0,0)$ that is a
simple integral representation involving a single integral, rather
than as an expression that must still be traced over all 3-momenta, as
in \cite{naro,cornwall,gavrilov}. Second, we use this explicit
expression to make a direct comparison with independent results from
the derivative expansion approximation
\cite{hallthesis,shovkovy98}. Third, we show how the real and
imaginary parts of the effective action are related by dispersion
relations, connecting perturbative and nonperturbative
expressions. Finally, we show how the uniform semiclassical
approximation \cite{brezin,balantekin88} fits into the resolvent
approach, obtaining a simple semiclassical expression for the QED
effective action in a general time dependent, but spatially uniform
$E$ field. This expression is remarkably similar to Schwinger's
``proper-time'' expression for the contant field case.

When the background field has constant field strength
$F_{\mu\nu}$, it is possible to obtain an explicit expression for the
exact effective action as an integral representation \cite{schwinger}. The
physical interpretation of this expression depends upon the magnetic
or electric character of the background, and this is reflected in how
we expand the integral representation.  In the case of a
constant magnetic background, a simple perturbative expansion in
powers of B yields
\begin{equation}
\Seff = {B^2 T L^3 \over 2 \pi^2} \sum_{n=1}^\infty
  {{\cal B}_{2n+2} \over (2n+2)(2n+1)(2n)} \left( 2B \over m^2
  \right)^{2n},
\label{constantmag}
\end{equation}
where the ${\cal B}_n$ are the Bernoulli numbers \cite{gradshteyn},
and $T L^3$ is the space-time volume factor.
In the case of a constant electric field background, the
 effective action is complex. The real part has a natural
 perturbative expansion which is just (\ref{constantmag})
 with $B\rightarrow iE$, while the imaginary part is a sum
 over nonperturbative tunnelling amplitudes
\begin{eqnarray}
{\rm Re} (\Seff) &=& -{E^2 T L^3 \over 2 \pi^2} \sum_{n=1}^\infty
  { (-1)^n {\cal B}_{2n+2} \over (2n+2)(2n+1)(2n)} \left( 2E \over
  m^2 \right)^{2n} \label{re part constant E-field}\\
{\rm Im} (\Seff) &=& {E^2TL^3 \over 8 \pi^3} \sum_{n=1}^\infty
  {1 \over n^2} e^{ - {m^2 \pi n \over E}}.
\label{im part constant E-field}
\end{eqnarray}

There are two clear motivations for studying the effective
action in non-constant background fields. First, knowledge
of the effective action for more general gauge fields is
necessary for the ultimate quantization of the electromagnetic
field. Second, realistic electromagnetic background fields do
not have constant field strength, and so we would like to
understand effective energies and pair production rates in
more general backgrounds. However, it is, of course, not
possible to evaluate the exact QED effective action for
a completely arbitrary background. Thus we are led naturally
to approximate expansion techniques. A common approach,
known as the derivative expansion
\cite{aitchison,cangemi95a,gusynin96,hallthesis},
involves a formal perturbative expansion about Schwinger's
exactly solvable case of constant field strength.
Unfortunately, this type of perturbative expansion is
difficult to
perform beyond first order, and is hard to interpret
physically, even for magnetic-type backgrounds. This is
even more problematic for electric-type backgrounds,
for which we seek a {\it nonperturbative} expansion.

A complementary approach is to search for solvable examples
that are more realistic than the constant field case, although
still not completely general. Recent work \cite{dunne98}
has found an exact, explicit integral representation for
the 3+1 dimensional QED effective action in a static but spatially
inhomogeneous magnetic field of the form
\begin{equation}
\vec{B}(\vec{x}) = ( 0,0,B \sech^2 \left( {x \over \lambda} \right)).
\label{magtanh}
\end{equation}
For fermions in this background field, there are three relevant
scales: a magnetic field scale $B$, a width parameter $\lambda$
characterizing the spatial inhomogeneity, and the fermion mass
$m$. It is therefore possible to
expand the exact effective action in terms of two independent
dimensionless ratios of these scales, depending on the
question of interest. For example, since $\lambda=\infty$
corresponds to the uniform background case, in order to
compare with the derivative expansion we expand the exact
$\Seff$ as a series in $1 \over B\lambda^2$. It has been
verified that the first two terms in this
series agree precisely with independent derivative
expansion results (there are no independent field theoretic
calculations of higher order terms in the derivative
expansion with which to compare). Furthermore, these and analogous
results in 2+1 dimensions indicate that the derivative expansion is in
fact an asymptotic series expansion \cite{cangemi95b,dunne97}.

Formally, one could change this magnetic-type result to an
electric-type background by an appropriate analytic
continuation $B
\rightarrow iE$. However, it is not immediately clear
how to obtain a {\it nonperturbative} expression [for
example, something like (\ref{im part constant E-field})]
for the imaginary part of the effective action. For constant
background fields a simple dispersion relation provides this
connection between the magnetic and electric cases, but for
nonconstant fields the
dispersion relations are more complicated. Understanding this
connection, for non-constant backgrounds, is one of the main
motivations for this paper.

This paper is organized as follows. In Section II we review
briefly the constant field case, using Schwinger's proper
time method. In Section III we review the resolvent method,
which has been used to obtain exact integral representations
for the effective action in the special nonuniform magnetic
 background (\ref{magtanh}). In Section IV we then use the
resolvent method to evaluate the exact effective action for
a time-dependent, but spatially uniform electric field
\begin{equation}
\vec{E}(\vec{x}) =( E \sech^2 \left( {t \over \tau} \right) ,0,0).
\label{electanh}
\end{equation}
In Section V we show how dispersion relations connect the
magnetic and electric cases (\ref{magtanh}) and (\ref{electanh}).
In Section VI we review the derivative expansion for electric fields
and in Section VII show its connection to the exact effective action
of Secion IV. In Section VIII we use a uniform semi-classical
approximation to obtain a general (but semi-classical) expression for
the pair production probability in a time-dependent electric
background. The final Section is devoted to some concluding comments.

\section{Schwinger's Approach}
Integrating over the fermion fields gives the QED effective
action for fermions in a background electromagnetic field
\begin{eqnarray}
\Seff[A] = - i \ln \det ( i\covar - m) = -{i \over 2} \tr\ln
 (\covar^2+m^2).
\label{schwingers trick}
\end{eqnarray}
Here, the covariant derivative is $\covar=\gamma^\mu
( \partial_\mu + iA_\mu)$ with the electric charge $e$ absorbed into
the gauge field $A$. In the calculations that follow we are implicitly
subtracting off zero field contribution $\Seff[A=0]$.

In a classic paper \cite{schwinger}, Schwinger computed the effective
action for constant background fields.  One expresses
the logarithm through an integral representation, the ``proper-time''
representation.
\begin{equation}
\Seff=-{i\over 2}\tr\ln(\covar^2+m^2)=  {i\over 2}
  \int_0^\infty{ds\over s} \tr e^{-s(\not{D}^2+m^2)}
\label{schwingers general}
\end{equation}
Clearly, to proceed, we need information concerning the spectrum
of the operator $\covar^2+m^2$.

For a constant magnetic background of strength B, we choose
$A_\mu=(0,0,0,By)$ and the Dirac representation of the gamma
matrices so that the operator becomes diagonal:
\begin{equation}
\covar^2+m^2= \left[\partial^2_0 -\partial_x^2-\partial_y^2
  -(\partial_z + iBy)^2 +m^2 \right] {\bf 1} +
  \left( \matrix { B&0&0&0 \cr 0&-B&0&0 \cr 0&0&B&0
  \cr 0&0&0&-B} \right).
\end{equation}
The Dirac trace is trivial, and we are left with a harmonic
oscillator system with eigenvalues
\begin{equation}
m^2-k_0^2+k_x^2 +2B(n+{1\over 2}\pm {1 \over 2})
\label{harmonic osc}
\end{equation}
The remaining traces are straightforward, yielding the exact
effective action for a constant magnetic field \cite{schwinger}
\begin{equation}
\Seff= {BTL^3 \over 8 \pi^2} \int_0^\infty {ds \over s^2}e^{-m^2s}
  \left( \coth Bs - {1 \over Bs} - {Bs \over 3} \right)
\label{integral constant B}
\end{equation}
Here, the ${1 \over Bs}$ term is an explicit subtraction of
$\Seff[0]$, while the  ${Bs \over 3}$ term corresponds
to a charge renormalization. A straightforward
expansion of (\ref{integral constant B}) yields the expansion
(\ref{constantmag}).

In a constant electric background the calculation is similar.
Choosing $A_\mu=(0,Ex_0,0,0)$ and using the chiral representation for
the gamma matrices, we find the operator $\covar^2+m^2$ diagonalizes:
\begin{equation}
\covar^2+m^2=[ \partial_0^2 - (\partial_x+iEt)^2 -\partial_y^2-
  \partial_z^2 +m^2] {\bf 1} +  \left( \matrix {
  iE&0&0&0 \cr 0&iE&0&0 \cr 0&0&-iE&0 \cr 0&0&0&-iE} \right).
\end{equation}
Once again, the Dirac trace is trivial, and we are left with a
harmonic oscillator with imaginary frequency. Thus
$\covar^2+m^2$ has complex eigenvalues
\begin{equation}
m^2 +2iE(n+{1\over 2}\pm {1 \over 2})+k_y^2+k_z^2
\label{imaginary ho}
\end{equation}
The traces can be performed as before, yielding
\begin{equation}
\Seff= {ETL^3 \over 8 \pi^2} \int_0^\infty {ds\over s^2} e^{-m^2s}
  \left(\cot Es -{1 \over Es} +{Es \over 3} \right)
\label{integral constant E}
\end{equation}
where we have subtracted the same vacuum contribution and charge
renormalization terms.

Going from (\ref{integral constant B}) to (\ref{integral constant E}),
we note poles of the integrand have moved onto the contour of
integration. This is the trademark of background electric fields and
the ultimate source of the imaginary contribution. Regulating the
poles with the standard principal parts prescription \cite{schwinger},
we separate out the imaginary and real contributions to the effective
action.
\begin{equation}
\Seff=i {E^2TL^3 \over 8\pi^2} \sum_{n=1}^\infty {1\over n^2}
  e^{-{m^2 \pi n \over E}} + {ETL^3\over 8 \pi^2} {\cal P}
  \int_0^\infty   {ds \over s^2} e^{-m^2s} \left( \coth Es -
  {1\over Es} -{Es\over 3} \right)
\end{equation}
As before, it is straightforward to expand the integral and arrive
at the expansion (\ref{re part constant E-field}), for the real part
of the effective action.

\section{Resolvent Method}
Now consider a class of more general backgrounds -- fields pointing
in a given direction and depending on only one space-time
coordinate. This is still far from the most general case;
nevertheless, this class is sufficiently broad to study the
effects of inhomogeneities, and yet simple enough to be
analytically tractable.

In the magnetic case we choose
\begin{equation}
\vec{A}= (0,0,a_B(y)) \hspace{1cm} \rightarrow \hspace{1cm}
  \vec{B}=(a^\prime_B(y),0,0)
\label{B general}
\end{equation}
while in the electric case we choose
\begin{equation}
\vec{A}= (a_E(t),0,0) \hspace{1cm} \rightarrow \hspace{1cm}
  \vec{E}=(a^\prime_E(t),0,0).
\label{E general}
\end{equation}

In the magnetic case there is no time dependence and $A_0=0$, so
we can perform the energy trace in (\ref{schwingers trick}).
After an integration by parts in $k_0$, this reduces the evaluation
of the effective action to a trace of a one-dimensional Green's
function, or resolvent
\begin{equation}
\Seff=-iL \int {dk_0\over 2\pi}  \sum_\pm \tr
  {k_0^2 \over {\cal D}_\pm(k_x,y,k_z) -k_0^2}
\label{B resolvent}
\end{equation}
where the one-dimensional operator ${\cal D}_\pm$ is
\begin{equation}
{\cal D}_\pm=m^2+k_x^2 -\partial_y^2 + (k_z-a_B(y))^2 \pm
  a^\prime_B(y).
\end{equation}

In the electric case there is no y dependence and $A_y=0$, so we
can perform the $k_y$ trace and obtain
\begin{equation}
\Seff=iL \int {dk_y \over 2\pi}  \sum_\pm \tr
  {k_y^2 \over {\cal D}_\pm(t,k_x,k_z) + k_y^2}
\label{E resolvent}
\end{equation}
which involves the resolvent of the operator
\begin{equation}
{\cal D}_\pm= m^2 + \partial_0^2 + (k_x - a_E(t))^2
  +  k_z^2 \pm ia_E^\prime(t).
\label{diff-e-q for electric}
\end{equation}

Thus, for both the magnetic and electric backgrounds in
(\ref{B general},\ref{E general}) the problem reduces to
tracing the diagonal resolvent (\ref{B resolvent},
\ref{E resolvent}) of a one-dimensional differential
operator. This makes clear the advantage of the resolvent
approach. For a typical background field we usually think
of computing the effective action by some sort of
summation over the spectrum of the appropriate Dirac operator.
This is easy for constant fields because the spectrum is
discrete [see (\ref{harmonic osc}) and (\ref{imaginary ho})].
But for non-constant fields the spectrum will typically
have both discrete and continuous parts, which makes a
direct summation extremely difficult. However, for
one-dimensional operators, we do not have to use this
eigenfunction expansion approach -- we can alternatively
express the resolvent as a product of two suitable independent
solutions, divided by their Wronskian. This provides a
simple and direct way to compute the effective action when
the background field has the form in (\ref{B general},
\ref{E general}).

This resolvent approach has been applied successfully to spatially
inhomogeneous magnetic backgrounds
\cite{cangemi95b,dunne97,dunne98,hallthesis}. It has also been used previously
by Chodos \cite{chodos} in an analysis of the possibility of spontaneous chiral
symmetry breaking for QED in time-varying background electric fields. In
this paper we present a detailed analysis of the resolvent approach to the
computation of the QED effective action in time-dependent electric backgrounds.
We
first check the resolvent approach by computing the effective action
for a constant electric field. The
constant electric field case follows the constant magnetic case
very closely. Choosing $a_E(t)=E t$, the eigenfunctions of the
operator (\ref{diff-e-q for electric}) are parabolic cylinder
functions. Taking independent solutions with the approriate
behavior at $t=\pm \infty$, we obtain the Green's function
\begin{equation}
{\cal{G}}(t,t')=-{\Gamma[-\nu] \over \sqrt{4 \pi iE}} D_\nu
  \left( \sqrt{ 2i \over E}  (Et-k_x)\right)  D_\nu \left(-
  \sqrt{2i \over E} (Et'-k_x)   \right)
\end{equation}
where we have defined $\nu={m^2 +k_y^2 +k_z^2 \over 2iE} \pm
{1\over2} + {1\over2}$. The trace of the diagonal Green's function
can be performed \cite{gradshteyn}, yielding Psi functions,
where $\psi(u)=\Gamma'(u)/\Gamma(u)$ is the logaritmic derivative
of the Gamma function \cite{gradshteyn}. Thus the effective action is
\begin{eqnarray}
\Seff & = & -{i L^3 \over 4 \pi^3} \int_{0}^{ET} dk_x
  \int_{-\infty}^{\infty}  k_y^2 dk_y dk_z  \sum_{\pm}
  \int_{-\infty}^{\infty} dx_0  {\cal{G}} (x_0,x_0) \nonumber \\
  & = & - {EL^3T \over 4 \pi^3} \int_{-\infty}^\infty k_y^2 dk_y dk_z
  \sum_{\pm} \left( \psi({1\over2}-{\nu\over2}) +\psi(-{\nu\over2})
  \right) \label{psi function} \\
  & = & {EL^3T \over 8 \pi^2} \int_0^\infty {ds \over s^2} e^{-m^2 s}
  \left( \cot Es - {1 \over Es} +{Es\over 3}\right).
\label{E integral 3}
\end{eqnarray}
The limits on the $k_x$ trace can be motivated by the classical
Lorentz interaction of the electron-positron pair after pair creation,
and can be checked by the requirement that the zero field part cancels
correctly. Note that the arguments of the psi functions appearing in
the effective action (\ref{psi function}) are complex. Thus we must
be careful to use the correct integral representation of the
$\psi$ function in the analysis. A convenient representation
for complex argument is given in Bateman \cite{bateman} as:
\begin{eqnarray}
\psi(z)&=& \log z -{1 \over 2z} -\int_0^{\infty e^{i\beta}} dt \left(
  {1\over e^t-1} -{1\over t} + {1\over2} \right) e^{-zt} \\
  & & -{\pi \over 2} < \beta < {\pi \over 2} \phantom{space}
  -\left({\pi \over 2} + \beta \right) < {\rm{arg}} z <
  \left( {\pi \over 2} + \beta \right) \nonumber
\end{eqnarray}
The expression (\ref{E integral 3}) is the same as
(\ref{integral constant E}), and
the calculation proceeds exactly as before. But for the constant field case
the resolvent method is unnecessarily complicated. The advantages of
the resolvent method will become evident when applied to more complicated
background fields, as is done in the remainder of this paper.

\section{Exactly Solvable Case}
In this section we apply the resolvent  method to a background
gauge field $A^\mu=(0,E \tau \tanh (\frac{t}{\tau}),0,0)$. This gauge
field corresponds to a single pulsed
electric field in the x-direction $E_x(t)=E \sech^2 (\frac{t}{\tau})$.
The electric field is spatially uniform but time-dependent; it
vanishes at $t=\pm\infty$, peaks at t=0, and has a temporal width
$\tau$ that is arbitrary. This field contains the constant field as a
special case when we take $\tau \rightarrow \infty$. The resolvent
expression (\ref{E resolvent}) for the effective action gives
\begin{equation}
\Seff =  i {L^3 \over 4\pi^3}  \int d^3k  \tr
  {k_y^2 \over \partial_0^2+ (k_x -{E \tau} \tanh ({t \over \tau}
  ))^2 +k_y^2 + k_z^2 +m^2 \pm i E \sech^2 \left({t \over \tau}\right)}
\end{equation}
The $k_x$ momentum trace runs over $(-\infty,\infty)$ since we
consider an infinite interaction time.

To determine the effective action we need the resolvent, which is
constructed from solutions to the ordinary differential equation
\begin{equation}
\left(\partial_0^2 +m^2+k_y^2 +k_z^2 +\left( k_x -E \tau
  \tanh\left({t \over \tau}\right) \right)^2 \pm iE \sech^2
  \left({t\over \tau}\right)  \right)  \phi=0
\end{equation}
This can be converted, by the substitution $y=\frac{1}{2}
(1+\tanh\frac{t}{\tau})$, to a hypergeometic equation,
with independent solutions
\begin{eqnarray}
\phi_1 & = & y^\alpha(1-y)^\beta \mbox{}_2F_1\left( {i \tau\over 2}
  \left( \alpha+\beta\pm 2 E\tau \right), {i\tau\over 2}
  \left(\alpha+\beta+1 \mp 2E\tau\right)  ; 1+ i \tau
  \alpha;y  \right) \nonumber \\
\phi_2 & = & y^\alpha(1-y)^\beta \mbox{}_2F_1\left( {i \tau\over 2}
  \left( \alpha+\beta\pm 2E\tau \right), {i\tau \over 2}
  \left(\alpha+\beta+1 \mp 2E\tau \right)  ; 1+ i\tau
  \beta;1-y  \right)
\end{eqnarray}
where we have defined
\begin{eqnarray}
y & = &{1\over2}(1+\tanh\left({t\over \tau}\right)) \nonumber \\
\alpha & = & \left( m^2+k_y^2
  +k_z^2 +\left(E\tau +k_x\right)^2 \right)^{1/2} \nonumber\\
\beta & = &  \left( m^2 +k_y^2 +k_z^2 + \left( E\tau
  -k_x \right)^2 \right)^{1/2}.
\label{a and b}
\end{eqnarray}
The boundary conditions are a particle of energy
$\alpha$ traveling forward in time and a particle of energy $-\beta$
traveling backward in time.

The diagonal resolvent is ${\cal G}(t,t)={\phi_1(t)\phi_2(t)\over
W[\phi_1,\phi_2]}$, where $W[\phi_1,\phi_2]$ is the Wronskian.
The trace over time once again yields psi functions (just as
in the magnetic cases treated in \cite{cangemi95b,dunne97}):
\begin{eqnarray}
\Seff & = & - {L^3 \tau \over 4\pi^3} \sum_\pm
  \int_{-\infty}^\infty {k_y^2 d^3k
  \over 4 } \left( {1\over\alpha} + {1\over\beta} \right)
  \left( \psi(1+ {i\tau\over2} ( \alpha + \beta \mp 2E \tau ))
  + \psi ( {i\tau\over 2} (
  \alpha + \beta \pm 2E \tau ) ) \right) \nonumber \\
  & = & - {L^3 \over 4\pi^3} \sum_\pm
  \int_{-\infty}^\infty {k_y^2 d^3k
  \over 4 k_\perp} {\partial \Omega_{(\pm)} \over \partial k_\perp}
  \left(\psi(1+{i \over 2} \Omega_\mp) + \psi({i\over2} \Omega_\pm)
  \right) \nonumber \\
  & = & {L^3 \over 4\pi^3} {1 \over2} \int d^3k \int_0^\infty
  {ds \over s} \left( e^{-\Omega_+ s} + e^{-\Omega_- s} \right)
  \left( \cot s -{1\over s} \right)
  \label{full action for tanh}
\end{eqnarray}
where we have defined $\Omega_+={\tau \over 2}\left( \alpha+\beta
+{2E\tau}\right)$ and $\Omega_-= {\tau \over 2 }
\left(\alpha+\beta-{2E\tau} \right) $.

Equation (\ref{full action for tanh}) is the exact effective action
for this time-dependent background gauge field. Notice the close
similarity to Schwinger's expression (\ref{E integral 3}) for the
constant background electric field. It is straightforward to
check that taking $\tau \rightarrow \infty$ reduces
(\ref{full action for tanh}) to the constant field result
(\ref{E integral 3}).

The effective action (\ref{full action for tanh}) has both real
and imaginary parts. As described before for the constant field
case, we regulate the integral using the principal part
prescription to obtain the imaginary part
\begin{eqnarray}
{\rm{Im}}(\Seff)& = & {1\over2} {L^3 \over 4\pi^3} \int d^3k\;
  \sum_{n=1}^\infty {1\over n} \left(e^{-n \pi \Omega_+} + e^{-n \pi
  \Omega_-} \right)  \nonumber \\
  & = & -{1 \over 2} {L^3 \over 4\pi^3} \int d^3k \;\ln\left(
  (1-  e^{-\pi \Omega_+} ) (1-e^{- \pi \Omega_-} ) \right)
\label{non expanded im part of tanh eff action}
\end{eqnarray}
and real part of the exact effective action
\begin{eqnarray}
{\rm{Re}}(\Seff) & = & {1 \over 2} {i L^3 \over 4\pi^3} \int d^3k
  \int_0^\infty {ds \over s} \left(e^{-i \Omega_+ s} + e^{- i
  \Omega_- s} \right)  (\coth s -{1\over s})   \nonumber \\
  & = & {1 \over 6} {L^3 \over 4\pi^3} \int d^3k {1 \over \Omega_+}
  + {L^3 \over (2 \pi)^3} \sum_{n=1}^\infty { (-1)^n {\cal{B}}_{2n+2}
  \over (2n+2)(2n+1) }  \int d^3k \left( {2 \over \Omega_+} \right)^{2n+1}
\end{eqnarray}
where we have asymptotically expanded the integral over $s$ in inverse
powers of $\Omega_+$.
The first term can be regulated and absorbed by renormalization. In
the second term the $k$ integrals can be done to yield the
integral representation
\begin{equation}
{\rm{Re}}(\Seff^{ren})=- {2L^3 \tau^3\over 3  \pi^2}
  \int_0^\infty {dt \over e^{2 \pi t} -1 } \left( { t- E\tau^2
  \over v_- } \left( {m^2\tau^2} - v_-^2
  \right)^{3/2} \sin^{-1}(\frac{v_-}{\tau m}) +
  (E \rightarrow -E) \right)
\label{real part from computation}
\end{equation}
where we have defined $v_-= (t^2 -2t E \tau^2)^{1/2}$. This integral
may be expanded as
\begin{eqnarray}
{\rm{Re}} \Seff^{ren}&=& - {L^3 \tau m^4 \over 8 \pi^{3/2}}
  \sum_{j=0}^\infty {1 \over \Gamma(j+1)} \left({1 \over 2E\tau^2}
  \right)^j \nonumber \\
  && \hspace{2.5cm} \times \sum_{k=1}^\infty
  {\Gamma(2k+j)\Gamma(2k+j-2)\over \Gamma(2k+1)\Gamma(2k+j+{1\over2})}
  (-1)^{k+j}  {\cal{B}}_{2k+2j}
  \left( {2E \over m^2} \right)^{2k+j}
\label{real full expansion of tanh action}
\end{eqnarray}

We now compare these results to previous analyses. The real part
of the effective action is exactly the same as the effective action
for the magnetic $\sech^2$ background case (see eqs. (10 and (18)
in \cite{dunne98}), with the replacements $B \rightarrow iE$ and
$\lambda T\rightarrow \tau L$. Thus, our naive expectation that
this simple analytic continuation from a magnetic to an electric
background is borne out. But in an electric background we are more
interested in the imaginary part, which does not have this type of
 perturbative expansion. Rather, it has the nonperturbative form
(\ref{non expanded im part of tanh eff action}). This explains how
it is possible to obtain both a perturbative and a nonperturbative
expression, for the real and imaginary parts respectively, from
the exact effective action (\ref{full action for tanh}).
Balantekin {\it et al.} \cite{balantekin91} have also computed
this imaginary part of the effective action for a $\sech^2$ electric
field. Our result
(\ref{non expanded im part of tanh eff action}) agrees with their
expression (see eqn (3.29) of \cite{balantekin91}), once theirs
is symmetrized in $E\to -E$, as it must be to satisfy Furry's
theorem.  This difference is not important for the imaginary part, but
it is crucial for the consistency of the dispersion relations which
relate the real and imaginary parts, as we show in the next section.

\section{Dispersion Relations}
In the previous section we found an expression for the exact effective
action for a particular background electric field. This effective
action has both real and imaginary parts. Given the real or imaginary
part of the effective action there exist dispersion relations which
relate the two. Here, we exploit the cuts in the electron self-energy
function to analytically continue it to the entire complex plane.
We shall show that there exist simple dispersion relations between
the real and imaginary parts of the effective action, both at the
perturbative level and also at the level of the general expression
(\ref{full action for tanh}).

\subsection{Perturbative Dispersion Relations}
Expand the imaginary part
(\ref{non expanded im part of tanh eff action})
of the effective action in powers of $E^2$
\begin{eqnarray}
{\rm{Im}} (\Seff) & = &  {1\over2} {L^3 \over 4\pi^3 }
  \sum_{n=0}^\infty{1\over n} \int d^3k\left( e^{-n \pi \Omega_+}
  + e^{-n \pi \Omega_-} \right) \\
  & = &  {L^3 \over 4\pi^3} \sum_{n=1}^\infty {1 \over n} \int
  d^3k \left[
  e^{- 2 \pi n \tau \sqrt{\mu^2+k_x^2} } + {E^2 \over 2}
  e^{- 2 \pi n \tau \sqrt{\mu^2+k_x^2} } \left( 4n^2 \pi^2\tau^4
  - {2n\pi\mu^3 \tau^3 \over (\mu^2+k_x^2)^{3/2}}
  \right) \right. \nonumber \\
  & \mbox{}&  \phantom{need space he}
  + E^4 e^{-2 \pi n \tau} \sqrt{\mu^2+
  k_x^2} \left( {2n^4\pi^4 \tau^8\over 3} - {2n^3\pi^3\mu^2 \tau^7
  \over (\mu^2+k_x^2)^{3/2} } + {n^2\pi^2\mu^4 \tau^6 \over
  2(\mu^2+k_x^2)^3} \right. \nonumber  \\
  & \mbox{}& \phantom{space needs go right here}
  + \left. \left. {\pi n\nu^2 \tau^5 \over 4
  (\mu^2+k_x^2)^{5/2}} - {5\pi n \mu^2 k_i^2 \tau^5 \over 4
  (\mu^2+k_x^2)^{7/2}} \right)+\dots \right]
\label{imaginary piece expansion in E^2}
\end{eqnarray}

Consider first the $E^2$ term. Doing the angular integrals we obtain
\begin{eqnarray}
[{\rm{Im}} \Seff]_{E^2}& = & {L^3 \over 4\pi^3} E^2 \sum_{n=1}^\infty
  {1 \over 2n} \int d^3k
  e^{- 2 \pi n \tau \sqrt{m^2+k^2} } 2 \pi n  \tau
  \left(2 n \pi \tau - { \tau(m^2+k^2 \sin^2 \theta) \over
  (m^2+k^2)^{3/2} } \right) \nonumber \\
  & = & {L^3 \over 4\pi^3}
  {4E^2 \pi^2 \tau^3\over 3 } \sum_{n=1}^\infty \int_0^\infty
  dk e^{- 2 \pi n \tau \sqrt{m^2+k^2}} \left( 6 \pi nk^2\tau
  - {3k^2m^2 + 2k^4 \over (m^2+k^2)^{3/2}} \right)
\end{eqnarray}
With the substitution $q=2 \sqrt{m^2+k^2}$ this becomes
\begin{eqnarray}
[{\rm{Im}} \Seff]_{E^2}& = &{L^3 \over 4\pi^3}
  {E^2 \pi^2 \tau^3 \over 3 } \sum_{n=1}^\infty
  \int_{2m}^\infty dq e^{n \pi q\over \lambda} (q^2-4m^2)^{1/2}
  \left( {3n \pi \tau} q - {2\over q^2}
  (q^2+2m^2) \right)\nonumber\\
& = & {L^3 \over 4\pi^3}
  {E^2 \pi^3 \tau^4 \over6} \int_{2m}^\infty dq
  q^2 \csch^2 {\pi q \tau\over 2} \left(1-{4m^2\over q^2}
  \right)^{1/2}  \left( 1+{2m^2\over q^2} \right) \nonumber \\
  \label{imaginary E^2 action}
  & = & {L^3 \over 4\pi^3} 4E^2 \pi^4 \tau^4
  \int_0^\infty dq q^2 \csch^2 {\pi q \tau \over 2 } {\rm{Im}}
  \Pi(q^2)
\end{eqnarray}
This expression agrees with the result of Itzykson and Zuber
\cite{itzykson}, where $\Pi(q^2)$ is the one-loop self-energy.
They reduced the problem to a one-dimensional Lippman-Schwinger
equation and expanded perturbatively to find the $E^2$ order term.

Along the real axis, there is a cut in the $q^2$ complex plane
from $(-\infty,-2m]$ and from $[2m,\infty)$. To derive the
dispersion relations we will need to consider an integral as $q^2
\rightarrow \infty$. Since the electron self-energy does not go to
zero as $q^2$, we need to add a linear convergence factor. The
convergence factor gives a residue at the origin which will ultimately
be absorbed by renormalization. This results in a once-subtracted
dispersion relation as follows.

Apply Cauchy's integral theorem to a function $f(z)$ satisfying these
properties. Let the contour be from $(-\infty,\infty)$ along the real
axis and close with an arc of infinite radius in the upper half plane.
\begin{eqnarray}
{f(z) \over z} & = & {1 \over 2\pi i} \oint_{\cal{C}}
  { f(\xi) d\xi \over \xi  (\xi -z)} \nonumber \\
  & = & {f(0) \over 2 z} + {1 \over 2\pi i} {\rm{P}}
  \int_{-\infty}^\infty {f(x^\prime) dx^\prime \over x^\prime
  (x^\prime -z)}
\end{eqnarray}
Now let the point z go to the real axis $z \rightarrow
x+i\varepsilon$.
\begin{equation}
\label{once subtracted dispersion}
{f(x) \over x} = {f(0) \over x} + {1 \over \pi i} {\rm{P}}
  \int_{-\infty}^\infty {f(x^\prime) dx^\prime \over x^\prime
  (x^\prime-x)}
\end{equation}
Take the real and imaginary parts of
(\ref{once subtracted dispersion}) and assume that $f(z)$ satisfies
the Schwarz reflection principle $f(z^*)=f^*(z)$.
\begin{eqnarray}
{\rm{Re}} (f(x)-f(0))
  &=&  {x\over \pi} {\rm{P}} \int_{-\infty}^\infty {{\rm{Im}}
  f(x^\prime) dx^\prime \over x^\prime (x^\prime -x) }
  =  {2 x^2 \over \pi} {\rm{P}} \int_0^\infty {{\rm{Im}} f(x^\prime)
  dx^\prime \over x^\prime ({x^\prime}^2 -x^2) } \\
{\rm{Im}} (f(x) -f(0))
  &=& -{x\over \pi} {\rm{P}} \int_{-\infty}^\infty {{\rm{Re}}
  f(x^\prime) dx^\prime \over x^\prime (x^\prime -x) }
  = -{2 x \over \pi}  {\rm{P}} \int_0^\infty {{\rm{Im}} f(x^\prime)
  dx^\prime \over x^\prime ({x^\prime}^2 -x^2) }
\end{eqnarray}
{}From the imaginary part of the electron self-energy in
({\ref{imaginary E^2 action}})
\begin{equation}
{\rm{Im}} \Pi(k) = {1\over 24 \pi} \left( 1-{4m^2 \over k^2}
  \right)^{1/2} \left( 1 +{2m^2 \over k^2} \right) \Theta(k^2 -4m^2)
\end{equation}
we can obtain the real part.
\begin{eqnarray}
{\rm{Re}} (\Pi(k) -\Pi(0)) & =& {2k^2 \over \pi} {1 \over 24 \pi}
  {\rm{P}} \int_{2m}^\infty {dk^\prime \over k^\prime ( {k^\prime}^2
  -k^2) } \left(1-{4m^2 \over {k^\prime}^2} \right)^{1/2} \left( 1 +
  {2m^2 \over {k^\prime}^2 } \right) \nonumber \\
  & = & {1 \over 32 \pi^{3/2}} \sum_{j=1}^\infty {\Gamma(j+2) \over j
  \Gamma(j+{5\over2}) } \left( {k\over 2m} \right)^{2j}
\end{eqnarray}
This is the kernel for the $E^2$ order real part of the effective
action.
\begin{eqnarray}
[{\rm{Re}} \Seff]_{E^2} & = & {L^3 \over 4\pi^3} 4E^2 \pi^4 \tau^4
  \int_0^\infty dq q^2 \csch^2 {\pi q \tau \over 2 }
  {\rm{Re}} ( \Pi(q^2) -\Pi(0) ) \nonumber \\
  & = & {L^2 \over 4\pi^3} {4 \pi^4 E^2 \tau^4 \over 32 \pi^{3/2}}
  \sum_{j=1}^\infty {\Gamma(j+2) \over j\Gamma(j+2)} {1\over
  (2m)^{2j}} \int_{-\infty}^\infty dq q^{2j+2}\csch^2
  {q\pi\tau\over 2}  \nonumber \\
  & = & {E^2 L^3 \tau\over 4 \pi^{3/2} } \sum_{j=1}^\infty
  {(-1)^j \Gamma(j+2) \over j\Gamma(j+{5\over2})} {\cal{B}}_{2j+2}
  \left({1 \over m\tau} \right)^{2j}
\label{dispersion real E^2 expansion}
\end{eqnarray}
This agrees with the $k=1$ term of
(\ref{real full expansion of tanh action}), the real part of the full
effective action to order $E^2$.

A similar analyis can be done for the $E^4$ contribution. Doing the
angular integrals in the $E^4$ piece from
(\ref{imaginary piece expansion in E^2}) gives
\begin{eqnarray}
[{\rm{Im}}\Seff]_{E^4} &=& {L^3 \over 4\pi^3} {4\pi^2 \tau^5\over 3}
  \sum_{n=1}^\infty \int_0^\infty k^2 dk e^{-2\pi n \tau
  \sqrt{m^2+k^2}} \left( 2n^3 \pi^3\tau^3 -
  {2n^2\pi^2\tau^2(2k^2+3m^2) \over (m^2+k^2)^{3/2}} \right.
  \nonumber \\
  & \mbox{} & \phantom{some space here} + \left.
  { n\pi \tau(15m^4+20m^2k^2+8k^4)\over 10(m^2+k^2)^3 } +
  {3 m^4 \over 4(m^2+k^2)^{7/2}} \right)
\end{eqnarray}
The substitution $q=2\sqrt{m^2+k^2}$ leads to
\begin{eqnarray}
[{\rm{Im}} \Seff]_{E^4} &=& {L^3 \over 4\pi^3} {\pi^2\tau^5 \over 3}
  \sum_{n=1}^\infty \int_{2m}^\infty dq q (q^2-4m^2)^{1/2} e^{-n\pi q
  \tau} \left( n^3 \pi^3\tau^3 -
  { 4n^2\pi^2\tau^2 (q^2+2m^2) \over q^3} \right. \nonumber \\
  & \mbox{} & \phantom{some space here} \left. +
  {8n\pi\tau(q^4+2m^2q^2+6m^4) \over 5 q^6} +
  {48m^4\over q^7} \right)
\end{eqnarray}
Integrate by parts in the $1^{st},2^{nd}$ and $4^{th}$ terms and
collect terms:
\begin{eqnarray}
[{\rm{Im}}\Seff]_{E^4} &=& -{L^3\over 4\pi^3} {8 \pi^3E^4m^4 \tau^6
  \over 3} \int_0^\infty dq d^4 \csch^2 {\pi q \tau\over 2}
  \Theta(q^2-4m^2)      \nonumber \\
  & \mbox{}& \phantom{some big space here} \times
  {1\over q^8} \left(1-{4m^2 \over q^2}\right)^{-3/2}
  \left( 3 -{10m^2 \over q^2} \right)
\end{eqnarray}
The dispersion relation for the $E^4$ term is derived in the same way
except that no subtraction is needed.
\begin{eqnarray}
{\rm{Re}}f(x) & = & {2\over \pi} {\rm{P}} \int_0^\infty {x^\prime
  dx^\prime \over {x^\prime}^2 -x^2} {\rm{Im}} f(x^\prime)
  \nonumber \\
{\rm{Im}}f(x) & = & - {2x\over \pi} {\rm{P}} \int_0^\infty {dx^\prime
  \over {x^\prime}^2 -x^2} {\rm{Re}}f(x^\prime)
\label{4th order dispersion relations}
\end{eqnarray}
With the dispersion relations (\ref{4th order dispersion relations})
we can immediately write down the complementary part of
the effective action at order $E^4$.
\begin{eqnarray}
[{\rm{Re}}\Seff]_{E^4} & = & - {L^3 \over 4\pi^3}
  {8 \pi^3E^4m^4 \tau^6\over 3}
  \int_0^\infty dq d^4 \csch^2 {\pi q\tau\over 2} {2\over\pi}
  {\rm{P}} \int_{2m}^\infty {k dk \over k^2-q^2} \nonumber \\
  & \mbox{} & \phantom{need some space here} \times
  {1 \over k^8}   \left( 1-{4m^2\over k^2} \right)^{-3/2}
  \left(3-{10 m^2 \over k^2} \right) \nonumber \\
  & = & - {2L^3E^4m^4 \tau^9\over \pi^{3/2}} \sum_{j=0}^\infty
  {(-1)^j \over \Gamma(j+1)} {\Gamma(j+4)\Gamma(j+2)\over
  \Gamma(5)\Gamma(j+{9\over2})} {\cal{B}}_{2j+4}
  \left( 1 \over m \tau \right)^{2j+8}
\label{dispersion E^4 term}
\end{eqnarray}
Thus, the dispersion relations have enabled us to deduce the $E^4$
term of real part (\ref{real full expansion of tanh action}) of the
effective action, beginning with the $E^4$ term in the imaginary part.

Using dispersion relations we have shown how it is possible to go from
a tunneling like expression to an asymptotic expansion at the first
two orders in $E^2$. Recall the real part for the exact effective
action with a $\sech^2$ background electric field
(\ref{real full expansion of tanh action}) is an asymptotic
expansion in {\it two} dimensionless scales ${1\over \tau^2}$ and
$\left({E \over m^2}\right)^2$. Following steps similar to those taken
above, we can find similar disperion relations for
the other expansion scale, ${1\over \tau^2}$. These relations have
been derived and are presented in \cite{hallthesis}.

\subsection{All-Orders Dispersion Relations}

The above approach could be continued to higher orders in $E^2$,
but the integrals become more difficult. Instead, we look for a
dispersion relation connecting the full exact expressions for
the real part (\ref{real part from computation}) and the imaginary
part (\ref{non expanded im part of tanh eff action}) of the
effective action. Begin with the imaginary part
(\ref{non expanded im part of tanh eff action}):
\begin{equation}
{\rm{Im}}\Seff={L^3 \over 4\pi^3}
  {1\over2} \sum_{n=1}^\infty {1\over n} \int d^3k \left(
  e^{-n \pi \tau \left(- 2E\tau +\sqrt{ \mu^2 +
  \left( E\tau +k_x\right)^2 } +\sqrt{ \mu^2+ \left(
  E\tau -k_x \right)^2} \right)} +(E \rightarrow -E) \right)
\label{full non-perturbative im part}
\end{equation}
Make the following substitution to unravel the exponents
\begin{equation}
2t=2E\tau^2 + \tau \sqrt{m^2 + E^2\tau^2 +k^2+ 2 E \tau k\cos\theta}
  +\tau \sqrt{ m^2 + E^2\tau^2 +k^2 -2 E\tau k  \cos\theta}.
\label{expression for t}
\end{equation}
Solve (\ref{expression for t}) for $k$,
\begin{equation}
k= \lambda \sqrt{ \left(t-E\tau^2\right)^2
  \left(t^2 - m^2 \tau^2 -2t E \tau^2 \right)
  \over t \left(t- 2E\tau^2 \right) - E^2 \tau^4 \sin^2 \theta }
\end{equation}
substitute into (\ref{full non-perturbative im part}), and do the
angular integration
\begin{eqnarray}
{\rm{Im}} \Seff &=& {L^3 \over 4\pi^3} \sum_{n=1}^\infty
  {\pi \over n} \int_{E\tau^2} \sqrt{m^2\tau^2 + E^2\tau^4}^\infty
  dt e^{-2\pi nt} {d \over dt} \int_0^{2\pi} d\theta
  \sin\theta \left(k^3(E) +k^3(-E) \right) \nonumber \\
  & = & -{L^3\over 4\pi^3}{4 \pi^2 m^4\tau \over 3} \int_0^\infty
  {dt \over e^{2\pi t} -1} \left(\Theta(z_--1) z_-^3 {dz_- \over dt}
  \left(1-{1\over z_-^2} \right)^{3/2} + (z_- \rightarrow z_+) \right)
\end{eqnarray}
where we have defined $z_-={1\over m \tau}(t^2 -2t E  \tau^2)^{1/2}$.

A dispersion relation can be derived for the complex variable
$z_-$. We regard the factor $\left(1-{1\over z_-^2}\right)^{3/2}$
as the imaginary part of an analytic
function defined along the whole real axis. Care must be taken since
the function doesn't go to zero along the arc as $z_- \rightarrow
\infty$; so we must insert a
convergence factor. There is a dispersion relation giving the real
part in terms of the imaginary part of a function with these
characteristics.
\begin{equation}
{\rm{Re}}(f(z_-)-f(0)) = {2 z_-^2 \over \pi} {\rm{P}} \int_0^\infty
  {{\rm{Im}} f(k) dk \over k (k^2-z_-^2) }
\label{dispersion for nonper}
\end{equation}
With (\ref{dispersion for nonper}) we can obtain the real part of the
effective action.
\begin{eqnarray}
{\rm{Re}}\Seff^{ren} & = & -{L^3\pi^2m^4\tau\over 3\pi} \int_0^\infty {dt
  \over e^{2\pi t} -1} \left( z_-^3 {dz_-\over dt} {2 z_-^2\over \pi}
  {\rm{P}} \int_1^\infty { \left( 1-{1\over k^2} \right)^{3/2} dk
  \over k(k^2-z_-^2) } + (z_- \rightarrow z_+) \right) \nonumber \\
  & = & -{m^4 L^3 \tau\over 15 \pi^2} \int_0^\infty {dt \over e^{2
  \pi t} -1} \left( z_-^4 {dz_-^2 \over dt}
  \mbox{}_2F_1(1,1;{7\over2},z_-^2) + (z_- \rightarrow z_+) \right)
  \label{has 2F1 in it}
\end{eqnarray}
In the last equation, we recognize the hypergeometric funciton
$\mbox{}_2F_1$, which has another representation in terms of
$\sin^{-1}$ \cite{gradshteyn}.
\begin{eqnarray}
{\rm{Re}}\Seff^{ren} & = & - {m^4 L^3 \tau\over \pi^2}
  \int_0^\infty {dt \over e^{2\pi t} -1} {2\lambda^2 \over m^2}
  \left( \left( t-{E \tau^2}\right) \right. \nonumber \\
  \phantom{some space here} & \times &
  \left. \left( -{2\over3} + {8  v_-^2 \over 9 m^2 \tau^2}
  + {2m \tau\over 3 v_-} \left( 1-{  v_-^2 \over
  \tau^2 m^2}\right)^{3/2} \sin^{-1} {v_-\over \tau m} \right)
  + (E \rightarrow  -E) \right)
\end{eqnarray}
We drop terms independent of E (since these cancel against the vacuum
subtraction)  and get
\begin{equation}
{\rm{Re}}\Seff^{ren} = -{2 L^3 \over 3 \pi^2\tau^3} \int_0^\infty
  {dt \over e^{2 \pi t} -1} \left( { t-E\tau^2
  \over v_-} \left( m^2\tau^2 -v_-^2 \right)^{3/2}
  \sin^{-1} {v_- \over \tau m} + (E \rightarrow -E) \right)
\label{total real from dispersion}
\end{equation}
where $v_-=\left( t^2 -2tE \tau^2 \right)^{1/2}$ and
$v_+=(t^2 + 2tE \tau^2 )^{1/2}$.
This expression for the real renormalized effective action is
exactly the same as obtained by a direct computation
(\ref{real part from computation}). Thus the dispersion
relations enable us to compute the real part, given the imaginary
part. The reverse direction works similarly.

\section{Derivative Expansion in 3+1 Dimensional Electric Field}
Schwinger solved the effective action exactly for constant background
fields. To solve for more realistic fields one must use some
perturbative expansion such as the derivative expansion. In the
derivative expansion the fields are assumed to vary very slowly. We
rewrite the trace in (\ref{schwingers general}) as a supersymmetric
quantum-mechanical path integral, expand the gauge field in a Taylor
series about the constant case, and interpret the successive
coefficients as successively increasing $n$-body interaction terms.
This has been done for 2+1 dimensional electric fields
\cite{hallthesis} and we may immediately generalize to 3+1 dimensions
by making the substitution $m^2 \rightarrow m^2+k_z^2$ and tracing
over the additional momentum \cite{dunne98}.
We obtain the zeroth and first orders of the derivative expansion for
a spatially homogeneous electric field in 3+1 dimensions
\begin{equation}
S= {i\over 2} \int d^4x \int_0^\infty {ds\over s}
  {e^{-m^2 s}\over 4i (\pi s)^2} \left[ (Es \cot Es -1)
  +(\partial_0 E)^2 \left({s^2\over 8 E^4} \right)
  (Es \cot Es)''' \right].
\label{full E deriv exp}
\end{equation}

Regulating the $s$ integral as before with the principal parts
prescription, we easily separate the real and imaginary parts of the
zeroth order derivative expansion term
\begin{eqnarray}
{\rm Im}[\Seff]_0 &=&   \int d^4x {E^2 \over 8\pi^3}
  \sum_{n=1}^\infty {1\over n^2} e^{-{m^2 \pi n \over E}}
\label{im zero order der exp} \\
{\rm Re}[\Seff]_0 &=&  \int d^4x {\cal P} \int_0^{ \infty}
  {ds \over s} {e^{-m^2 s}\over 8 \pi^2 s^2} \left( Es \cot Es
  -1 \right).
\label{re zero order der exp}
\end{eqnarray}
Perform an asymptotic expansion of the integral over $s$
and we obtain
\begin{equation}
{\rm Re}[\Seff]_0= -\int d^4x {E^2 \over 2 \pi^2}
  \sum_{n=1}^\infty { (-1)^n {\cal B}_{2n+2}
  \over 2n(2n+1)(2n+2)} \left( 2 E \over m^2
  \right)^{2n}
\label{re zero order der exp 2}
\end{equation}
where ${\cal B}_{\nu}$ is the $\nu^{th}$ Bernoulli number.
Equations (\ref{im zero order der exp}) and
(\ref{re zero order der exp 2}) are the same as the corresponding
equations for the constant field result (\ref{im part constant E-field})
and (\ref{re part constant E-field}), with the constant field $E$ replaced
by the time dependent field $E(t)$.

Now consider first derivative term in (\ref{full E deriv exp}).
Separating out the imaginary component is complicated by the fact that
the triple derivative introduces fourth order poles along the real
axis, while in the zeroth order term the poles are of first
order. The exact effective action, containing both imaginary and
real components, for the first order derivative term is
\begin{eqnarray}
[\Seff]_1&=&{i \over 2} \int d^4x \int_0^\infty {ds \over s} {e^{-m^2 s}
  \over 4i (\pi s)^2} (\partial_0 E)^2 {s^2 \over 8 E^4}
  (s E \cot sE -1)''' \nonumber   \\
&=& - {1 \over 64 \pi^2} \int d^4x { (\partial_0 E)^2 \over
  E^4} \sum_{n=1}^\infty \int_0^\infty {ds \over s}
  {e^{-m^2 s} \over E^4 \left(s-{n \pi \over E}\right)^4 }
  {48 n^2 \pi^2 E^4 s (n^2 \pi^2 +s^2 E^2)
  \over(Es+ \pi n)^4 }.
\end{eqnarray}
In this expression we clearly see the presence of the fourth order
poles along the real axis. Regulating using the principal parts
prescription, we get the imaginary part which is just a sum of
${1\over2}$ the residues
\begin{eqnarray}
{\rm Im}[\Seff]_1 &=&
  -{1\over 64 \pi^2} \int d^4x { (\partial_0 E)^2
  \over E^4} \sum_{n=1}^\infty {\pi \over 3!} \left.\left( {e^{-m^2 s}
  \over s}  {48n^2 \pi^2 s(n^2 \pi^2+s^2E^2)\over (Es+\pi n)^4}
  \right)''' \right|_{s \rightarrow {\pi n \over E}} \nonumber \\
&=& {1 \over 64 \pi} \int d^4x {(\partial_0 E)^2 \over E^4}
  \sum_{n=1}^\infty { e^{-{n m^2 \pi \over E}} \over (\pi n)^3}
  (6E^3+6E^2m^2n\pi+3Em^4n^2\pi^2+m^6n^3\pi^3).
\label{im first order der exp}
\end{eqnarray}
As before asymptotically expand the integral in powers of
${E\over m^2}$ and we find
\begin{equation}
{\rm Re}[\Seff]_1={m^6 \over 64 \pi^2}\int d^4x {(\partial_0 E)^2
  \over E^4} \sum_{n=1}^\infty { (-1)^n {\cal B}_{2n+2}
  \over 2n-1} \left( {2E \over m^2} \right)^{2n+2}.
\label{re first order der exp}
\end{equation}
Note that in the spirit of the derivative expansion approximation, $E$
means $E(t)$ in the expressions
(\ref{full E deriv exp}-\ref{re first order der exp}). In the next
section we will specialize to the $\sech^2$ electric field and compare
with the exact result (\ref{full action for tanh}) for the effective
action.

\section{Derivative Expansion in Exactly Solvable Case}
For the electric field
\begin{equation}
E_1(t)=E \sech^2 \left({t \over \tau}\right).
\label{another sech}
\end{equation}
the exact effective action is (\ref{full action for tanh}), with
explicit real and imaginary parts in
(\ref{real full expansion of tanh action}) and
(\ref{non expanded im part of tanh eff action}), respectively. In
order to compare with the derivative expansion results in
(\ref{im zero order der exp}), (\ref{re zero order der exp 2}),
(\ref{im first order der exp}) and (\ref{re first order der exp}), we
still need to perform the $t$ integrals in these expressions, with
$E(t)=E \sech^2 \left({t\over \tau}\right)$.

\subsection{Comparison of real part}
Insert the electric field (\ref{another sech}) into the real part of
the zero order derivative expansion effective action
(\ref{re zero order der exp}) and do the $t$ integral using the
formula (3.512.2) from Gradshteyn \cite{gradshteyn}
\begin{equation}
\int_0^\infty {\sinh^\mu x \over \cosh^\nu x}dx=
{ \Gamma({\mu+1\over2}) \Gamma({\nu -\mu \over2}) \over 2\Gamma({\nu +1
\over 2})}
\end{equation}
and we obtain
\begin{equation}
{\rm Re}[\Seff]_0=-{\tau L^3 m^4 \over 8\pi^{3/2}} \sum_{n=1}^\infty
  {\Gamma(2n-2) \Gamma(2n) \over \Gamma(2n+1) \Gamma(2n+{1\over2}) }
  (-1)^n {\cal B}_{2n} \left( {2E \over m^2} \right)^{2n}.
\label{sech j=0}
\end{equation}
This is precisely the leading term, as an expansion in ${1\over
E\tau^2}$, of the exact effective action
(\ref{real full expansion of tanh action}). Similarly, for the real
part of the first order derivative term
(\ref{re first order der exp}) in the expansion of the of the
effective action, doing the $t$ integral yields
\begin{equation}
{\rm Re}[\Seff]_1={L^3 m^2\over 8\pi^{3/2}\tau} \sum_{n=1}^\infty
  { \Gamma(2n+1) \Gamma(2n-1) \over \Gamma(2n+1) \Gamma(2n+{3\over2})}
  (-1)^n {\cal B}_{2n+2} \left({2E\over m^2}\right)^{2n}.
\label{sech j=1}
\end{equation}
This is precisely the next-to-leading term in the expansion
(\ref{real full expansion of tanh action}) of the exact result.

This agreement is as expected for the field (\ref{another sech}),
each order in the derivative expansion introduces an extra factor of
${1\over \tau^2}$. These results provide strong evidence that the
expansion (\ref{real full expansion of tanh action}) of the exact
result is an all-orders derivative expansion, as in the magnetic case
\cite{cangemi95b,dunne97}. However, as in the magnetic case, we note
that this is an asymptotic expansion.

\subsection{Comparison of imaginary part}
For the imaginary piece we follow a different approach to make
the comparison. Inserting the $E(t)=\sech^2\left({t\over \tau}\right)$
into the zero-order and first-order expressions
(\ref{im zero order der exp}) and (\ref{im first order der exp})
leads to the probability integral, which cannot be computed
explicitly. Instead, we expand the imaginary part of the exact
effective action (\ref{non expanded im part of tanh eff action}) in
inverse powers of $\tau$, and transform the momentum integrals into a
form which can be compared directly with the derivative expansion
answers (\ref{im zero order der exp}) and
(\ref{im first order der exp}).

Recall the imaginary part
(\ref{non expanded im part of tanh eff action})
of the exact effective action
\begin{equation}
{\rm Im}(\Seff)= {L^3 \over 4\pi^3} {1\over 2}\int d^3k
  \sum_{n=1}^\infty {1\over n} \left( e^{-n\pi\Omega_+} +
  e^{-n\pi\Omega_-} \right)
\end{equation}
where $\Omega_+$ and $\Omega_-$ are defined as
\begin{eqnarray}
\Omega_+=\tau(\alpha+\beta+2E\tau) \hspace{1cm}
\Omega_-=\tau(\alpha+\beta-2E\tau)
\end{eqnarray}
and $\alpha$ and $\beta$ are defined in (\ref{a and b}).
We can ignore the $\Omega_+$ term in the derivative expansion,
$\tau \rightarrow \infty$, since it
is suppressed by an exponential factor $e^{-4E\tau}$ relative to the
$\Omega_-$ piece. Make the transformation
\begin{equation}
2t=\tau\left(-2E\tau+\sqrt{\mu^2+(E\tau+k_x^2)^2}+
  \sqrt{\mu^2+(E\tau-k_x)^2}\right)
\end{equation}
and solve for $k_x$
\begin{equation}
k_x={(t+\tau^2 E) \sqrt{t^2-\mu^2\tau^2+2t\tau^2E} \over t^{1/2} \tau
  \sqrt{t+2\tau^2 E}}.
\end{equation}
The integral is now
\begin{equation}
{\rm Im}(\Seff)={L^3 \over 2\pi^2} \sum_{n=1}^\infty \int dk_y dk_z
  \int_{t_0}^\infty e^{-2\pi n t} k_x(t)
\end{equation}
where the lower limit on the integration is
$t_0=-\tau^2 E + \tau\sqrt{\mu^2+E^2 \tau^2}$. Make another
transformation to the coordinate $z$
\begin{equation}
z={1\over \mu\tau}\sqrt{t^2+2t\tau^2 E} \hspace{1cm}
{dt \over dz} ={\tau z \mu^2 \over \sqrt{\mu^2 z^2 +\tau^2 E^2}}
\end{equation}
and the integral becomes
\begin{equation}
{\rm Im}(\Seff)= {L^3\tau\over 2\pi^2} \sum_{n=1}^\infty
  \int dk_y dk_z  \mu^2  \int_1^\infty dz \sqrt{z^2-1}
  e^{-2\pi n\left(-\tau^2 E + \tau\sqrt{\mu^2 z^2 +\tau^2 E^2}
  \right)}
\end{equation}
which can be expanded in inverse powers of $\tau$.
\begin{eqnarray}
{\rm Im}(\Seff)&=&{L^3\tau\over 2\pi^2}\sum_{n=1}^\infty\int dk_y
  dk_z \mu^2 \int_1^\infty dz \sqrt{z^2-1} e^{-{n\pi z^2 \mu^2 \over E}}
  \left(1+{n\pi z^4 \mu^4 \over 4E^3\tau^2} +
  \ldots \right)
\label{expansion of im part}
\end{eqnarray}

Complete the integral over $k_y$ and $k_z$ in the leading term
\begin{eqnarray}
{\rm Im}[\Seff]_0&=&{L^3\tau E^2\over 4\pi^3}
  \sum_{n=1}^\infty {1\over n^2} \int_1^\infty {dz\over z^4
  \sqrt{z^2-1}}e^{-{n\pi m^2 \over E} z^2} \label{extract time dep} \\
&=&{L^3 \tau E^2 \over 8 \pi^3} \sum_{n=1}^\infty
  {1\over n^2} e^{-{n\pi m^2 \over E}} \Psi({1\over2},-1,
  {n\pi m^2 \over E}) \label{compare with gitman}
\end{eqnarray}
where $\Psi$ is the confluent hypergeometric function defined in
6.5(2) of Bateman \cite{bateman}. Gavrilov and Gitman \cite{gavrilov}
have found, by other methods, the zeroth order
term for this field configuration and obtain precisely
(\ref{compare with gitman}).
In order to compare with the zeroth order derivative expansion result
(\ref{im zero order der exp}) we substitute
$z=\cosh\left({t\over \tau}\right)$
in (\ref{extract time dep}) to obtain
\begin{equation}
{\rm Im}(\Seff)_\tau= \int d^4x {E^2(t)\over 8\pi^3} \sum_{n=1}^\infty
 {1\over n^2}  e^{-{n\pi m^2 \over E(t)}}
\end{equation}
where $E(t)=E\sech^2\left({t\over \tau}\right)$.  This is precisely
the imaginary part of the zeroth order term of the derivative
expansion (\ref{im zero order der exp}).

Similarly, perform the integrals over $k_y$ and $k_z$
in the next-to-leading order term in (\ref{expansion of im part})
\begin{eqnarray}
{\rm Im}[\Seff]_1&=& {L^3\over 8\pi \tau E^2}
  \sum_{n=1}^\infty  {1\over \pi^3 n^3} \int_1^\infty dz
  {\sqrt{z^2-1} \over z^4} e^{-{n\pi m^2 \over E}z^2}
  \nonumber \\ && \hspace{1cm} \times
  (6E^3+6 E^2m^2 n \pi z^2+3Em^4n^2 \pi^2z^4+m^6 n^3\pi^3z^6).
\end{eqnarray}
To compare with the first order derivative expansion result
(\ref{im first order der exp}), we make the same substitution
$z=\cosh \left({t\over\tau}\right)$ to obtain
\begin{eqnarray}
{\rm Im}[\Seff]_1&=&{1\over64\pi}\int d^4x
  {(\partial_0E)^2\over E^4}   \sum_{n=1}^\infty
  {1\over n^3\pi^3} e^{-{n\pi m^2 \over E}}
  \nonumber \\ && \hspace{1cm} \times
  (6E^3+6E^2 m^2 n\pi +3E^3m^4 n^2\pi^2 +m^6n^3\pi^2).
\end{eqnarray}
where here $E$ means $E(t)=E \sech^2 \left({t\over \tau}\right)$.
This is the same result we obtained for the first derivative term of
the imaginary part of the effective action
(\ref{im first order der exp}).  As with the real part of the
effective action, successive terms in inverse powers of $\tau$ from
(\ref{expansion of im part}), correspond to increasing orders of the
derivative expansion.

\section{Exact Semi-Classical Action for More General Fields}
As discussed in Section III, the resolvent method is a useful
technique for evaluating the exact effective action when the Dirac
operator can be reduced to an effectively one-dimensional operator.
In this section we show how a generalized WKB expansion can then be
used to obtain an exact semi-classical
effective action for background electric fields with more general
time dependence than the $E(t)=E\sech^2\left({t\over \tau}\right)$
example considered in the previous two sections.

Assume the background gauge field has only one component in the
x-direction $A_\mu=(0,a(t),0,0)$. According to (\ref{E resolvent}),
we seek the Green's functions
\begin{equation}
-\left( \hbar^2 \partial_0^2 +\mu^2 + \phi^2(t) \pm i
  \hbar \phi^\prime(t) \right) \: {\cal{G}}^\pm_{k_\perp}(t,t^\prime)
  = \delta(t-t^\prime)
\label{diff e q for general case}
\end{equation}
where $\mu^2=m^2 +k_y^2 + k_z^2$, and $\phi=a(t)-k_x$.

In the uniform semiclassical approximation \cite{balantekin88},
one begins by looking for solutions $\psi(t)=K(t)
U\left(S(t)\right)$.
The familiar WKB approximation of quantum mechanics consists of
the choice: $\psi(t)=K
e^{iS(t)}$. Instead, a uniform semiclassical approximation is
obtained by choosing $U$ to be a parabolic cylinder function.
Define $U$ to satisfy
\begin{equation}
- \hbar^2 {\partial^2 U \over \partial S^2} - (S^2 + i \eta \hbar)
  U(S) = \Omega U(S)
\end{equation}
to which independent solutions are
\begin{equation}
D_\nu \left( \pm {1+i \over \sqrt{\hbar} } S(t) \right)
\end{equation}
where $\eta$ goes as the sign of $\phi^\prime$, and $\nu={1\over2} (\eta
-1-{i \over \hbar} \Omega)$. Now take
$K=(S^\prime)^{-1/2}$. Then the general differential equation
(\ref{diff e q for general case}) becomes a differential equation
relating $K$ and $S$.
\begin{equation}
\hbar^2 {1 \over K} {\partial^2K\over \partial t^2} - \left(
  {\partial^2 S \over \partial t^2} \right) (\Omega + i \eta \hbar +
  S^2) + (\mu^2 + \phi^2) \pm i \hbar \phi^\prime = 0
\end{equation}
Expand $S(t) \approx S_0(t)+\hbar S_1(t)$ and
collect the zeroth order terms in $\hbar$.
\begin{equation}
\mu^2 + \phi^2(t)=(\Omega +S_0^2) \left( {\partial S_0 \over \partial
    t} \right)^2
\label{zeroth order eq}
\end{equation}
The WKB expansion is a good approximation when the zeroth order term
outsizes the first order term $1 \gg | S_1/S_0 |$. At points $t^\prime$
where $S_0(t^\prime) \rightarrow 0$ the approximation doesn't work
unless we require $S_1(t^\prime) \rightarrow 0$ as well. Then apply
L'H\^{o}pital's rule
\begin{equation}
1 \gg  \left| { S_1 \over S_0 } \right| = \left| {S_1^\prime  \over
    S_0^\prime} \right| = \left| S_1^\prime \right| \left| \left(
    {\Omega + S_0^2 \over \mu^2 + \phi^2 } \right)^{1/2} \right|
\end{equation}
and we see the generalized WKB will be an appropriate expansion if the
turning points of the numerator $S_0(t_0)=i \sqrt{\Omega}$ and
$S_0(t_0^*)= -i\sqrt{\Omega}$ are the same as those of the denominator
$\phi(t_0)=i \mu$ and $\phi(t_0^*)=-i\mu$. Using the turning points,
we can integrate (\ref{zeroth order eq}) and find the quantity
$\Omega$ \cite{balantekin88}:
\begin{equation}
\int_{t_0}^{t_0^*} dt \sqrt{\mu^2+\phi^2(t)} =
\int_{t_0}^{t_0^*} dt {dS_0\over dt} \sqrt{\Omega+S_0^2} =
\int_{i\sqrt{\Omega}}^{-i\sqrt{\Omega}} dS_0 \sqrt{\Omega+S_0^2}=
- {i \Omega \pi \over 2}.
\label{omeg}
\end{equation}

Given the wavefunctions, we can express the Green's function as
\begin{equation}
{\cal{G}}_{k_\perp}^{\pm(\eta)}(t,t^\prime) = - {\Gamma(-\nu) \over 2
  \sqrt{\pi}} e^{-{i
  \pi \over 4}} {1 \over S^\prime} D_\nu \left( {1+i \over \sqrt{\hbar}}
  S(t) \right) D_\nu \left( - {1+i \over \sqrt{\hbar}} S(t^\prime) \right)
\end{equation}
The resolvent approach then gives the effective action as
\begin{equation}
\Seff= i { L^3 \over 4 \pi^3} \int k_y^2 d^3k {1\over 2} \sum_{\pm E}
  2 \sum_{\pm(\eta)} \int_{-\infty}^{\infty} dx_0\; \mbox{}
  {\cal{G}}_{k_\perp}^{\pm(\eta)}(t,t)
\end{equation}
where we explicitly summed signs of the electric field to satisfy
Furry's theorem.

Now make the semiclassical approximation by replacing $S(t)$ by
$S_0(t)$. The semiclassical Green's function is
\begin{equation}
{\cal{G}}_{k_\perp}^{\pm(\eta),sc} (t,t^\prime)=- { \Gamma(-\nu)
  \over 4\sqrt{\pi} }
  e^{- {i\pi \over4}} {1  \over k_\perp}
  {\partial \Omega \over \partial k_\perp}   S_0^\prime
  D_\nu \left( {1+i\over \sqrt{\hbar}} S_0(t) \right)
  D_\nu \left(-{1+i\over \sqrt{\hbar}} S_0(t^\prime) \right).
\end{equation}
where we have used the identity [see Eq. \ref{zeroth order eq}] that
$\frac{1}{S_0^\prime}=\frac{1}{2k_\perp} \frac{\partial \Omega}{\partial
k_\perp} S_0^\prime$.
The $t$ integral in the trace of the diagonal resolvent can be converted to an
integral over $S_0$, giving
\begin{eqnarray}
\Seff^{sc}& = & - {L^3 \over (2\pi)^3} \int_{-\infty}^\infty k_y^2
  d^3k \sum_{\pm E}
  {1 \over 2 k_\perp} { \partial \Omega \over \partial
  k_\perp} \left( \Psi({i\over2} \Omega)
  + \Psi( 1 + {i\over2} \Omega)  \right) \nonumber \\
  & = & {L^3 \over (2\pi)^3 } {1\over 2} \int d^3k \int_0^\infty
  {ds \over s } \left( e^{-\Omega(E) s}+ e^{-\Omega(-E) s}\right)
  (\cot s-{1\over s})
\label{main result}
\end{eqnarray}
This expression is the exact (but semiclassical) effective action
for an electric background field that is spatially uniform, but
 has general time dependence $\phi^\prime (t)$. The function
$\Omega$ is given by (\ref{omeg}).
It is interesting to note how similar this general expression
is to Schwinger's exact expression (\ref{integral constant E})
for the constant background field case.

In the exactly solvable case studied in the previous two sections,
$\phi(t)=E\tau\; \tanh(\frac{t}{\tau})$. In this case the integral
(\ref{omeg}) for $\Omega$ can be done exactly and we arrive at the
exact expression (\ref{full action for tanh}) derived before with the
resolvent method. The fact that the uniform
semiclassical approximation is actually exact in this case is due
to the supersymmetry underlying the uniform semiclassical
approximation in this system. In general cases that are not exactly
solvable, the expression (\ref{main result}) still gives the
semiclassical answer. For example, a periodic background gauge field
\mbox{$A_\mu=(0,{E \over \omega_0} \cos(\omega_0 t),0,0)$} is not
an exactly solvable case. However, the expression (\ref{main result})
immediately gives the semiclassical result of Brezin and Itzykson
\cite{brezin} [see (eq.44) of their paper] for the imaginary
part of the effective action in an alternating electric field.

\section{Conclusions}
In conclusion, we have used the resolvent approach to compute the
exact QED effective action for the time dependent electric field
background $\vec{E}(t)=( E\sech^2 \left({t\over
\tau}\right),0,0)$. The result is a simple integral representation
involving a single integral, just as in Schwinger's proper-time result
for the constant electric field case. We then used this exact result
to investigate the dispersion relations relating the real and
imaginary parts of the effective action. This explains the connection
between the nonperturbative form of the imaginary part, and the
perturbative form of the real part. It is this perturbative real part
that should be compared with results for magnetic backgrounds. In
addition, we made an asymptotic expansion of the exact answer in
powers of ${1\over E\tau^2}$, and showed that the first two terms
agree with (independent) results from the derivative
expansion. Finally, we showed how the uniform semiclassical approach
of Balantekin {\it el al.} is incorporated into the resolvent
approach, yielding a simple semiclassical expression that encodes both
the real and imaginary parts of the effective action.
The challenge now is to use these results for the effective action to
obtain realistic estimates of pair-production rates in electric fields
with practically attainable strength and time dependence.

\vspace{1cm}
\noindent{\bf Acknowledgments}

This work has been supported by the Department of Energy grant
No. DE-FG02-92ER40716.00., and the University of Connecticut Research
Foundation. We also thank Carl Bender and Alain Comtet for helpful
comments and suggestions.


\begin{thebibliography}{10}

\bibitem{schwinger}
J. Schwinger, Phys. Rev. {\bf 82},  664  (1951).

\bibitem{stone}
M. Stone, Phys. Rev. D {\bf 14},  3568  (1976).

\bibitem{brezin}
E. Brezin and C. Itzykson, Phys. Rev. D {\bf 2},  1191  (1970).

\bibitem{naro}
N. Narozhny\u{i} and A. Nikishov, Sov. J. Nucl. Phys. {\bf 11},  596  (1970).

\bibitem{cornwall}
J. Cornwall and G. Tiktopoulos, Phys. Rev. D {\bf 39},  334  (1989).

\bibitem{balantekin88}
A. Balantekin, S. Fricke, and P. Hatchell, Phys. Rev. D {\bf 38},  935  (1988).

\bibitem{balantekin91}
A. Balantekin, J. Seger and S. Fricke, Intl. J. of Mod. Phys. A {\bf 6},  695
(1991).

\bibitem{chodos}
A. Chodos, in {\it Proceedings of the G\"ursey Memorial Conference on Strings
and Symmetries} (Istanbul 1994), G. Aktas et al, Eds., {\it Lecture Notes in
Physics} Vol. 447 (Springer-Verlag, 1995).

\bibitem{gavrilov}
S. Gavrilov and D. Gitman, Phys. Rev. D {\bf 53},  7162  (1996).

\bibitem{burke}
D. Burke, Phys. Rev. Lett. {\bf 79},  1626  (1997).

\bibitem{melissinos}
A. Melissinos, hep-ph/9805507 (unpublished).

\bibitem{hallthesis}
T. Hall, Ph.D. thesis, University of Connecticut, 1998.

\bibitem{shovkovy98}
V. Gusynin and I. Shovkovy, hep-th/9804143 v3 (unpublished).

\bibitem{gradshteyn}
I. Gradshteyn and I. Rhyzhik, {\em Table of Integrals Series and Products}
  (Academic Press, New York, 1980).

\bibitem{aitchison}
I. Aitchison and C. Fraser, Phys. Rev. D {\bf 31},  2605  (1985).

\bibitem{cangemi95a}
D. Cangemi, E. D'Hoker, and G. Dunne, Phys. Rev. D {\bf 51},  2513  (1995).

\bibitem{gusynin96}
V. Gusynin and I. Shovkovy, Can. J. Phys. {\bf 74},  282  (1996).

\bibitem{dunne98}
G. Dunne and T. Hall, Phys. Lett. B {\bf 419},  322  (1998).

\bibitem{cangemi95b}
D. Cangemi, E. D'Hoker, and G. Dunne, Phys. Rev. D {\bf 52},  3163  (1995).

\bibitem{dunne97}
G. Dunne, Int. J. Mod. Phys. A {\bf 12},  1143  (1997).

\bibitem{bateman}
A. Erd\'{e}yli, {\em Higher Transcendental Functions} (Robert E. Kreiger
  Publishing Company, Malabar, FL., 1953).

\bibitem{itzykson}
C. Itzykson and J. Zuber, {\em Quantum Field Theory} (McGraw-Hill, New York,
  1980).

\end{thebibliography}

\end{document}